# Morphology-Preserving Holotomography: Quantitative Analysis of 3D Organoid Dynamics

ChulMin Oh, Jimin Cho, Juyeon Park, Hoyeon Lee, YongKeun Park

***Abstract*—Organoids are three-dimensional (3D) in vitro models for studying tissue development, disease progression, and physiological responses. Holotomography (HT) enables long-term, label-free imaging of live organoids by reconstructing volumetric refractive-index (RI) maps, but quantitative analysis is limited by the missing-cone artifact, which introduces anisotropic resolution and axial distortion. Here, we present a quantitative analysis framework that addresses the missing-cone problem at the level of image representation rather than reconstruction. We introduce morphology-preserving holotomography (MP-HT), a torus-shaped spatial filtering strategy that emphasizes high-spatial-frequency RI texture while suppressing low-frequency components most susceptible to missing-cone–induced distortion. Based on MP-HT, we develop a 3D segmentation pipeline for robust separation of epithelial and luminal structures, together with a model-based RI quantification approach that incorporates the system point spread function to enable morphology-independent estimation of dry-mass density and total dry mass. We apply the framework to long-term imaging of live hepatic organoids undergoing expansion, collapse, and fusion. The results demonstrate consistent segmentation across diverse geometries and reveal coordinated epithelial–lumen remodeling, breakdown of morphometric homeostasis during collapse, and transient biophysical fluctuations during fusion. Overall, this work establishes a physically transparent and reproducible approach for quantitative, label-free analysis of organoid dynamics in three dimensions.**

## I. INTRODUCTION

Organoids are three-dimensional (3D) multicellular systems that recapitulate key aspects of tissue architecture, polarity, and function beyond what is achievable in conventional two-dimensional cultures [1]. By self-organizing into epithelial structures with well-defined luminal compartments, organoids provide powerful in vitro models for studying development, disease progression, and drug responses at the tissue level [2]–[8]. Because organoid morphology and internal organization evolve dynamically over time [9]–[15], quantitative analysis of 3D morphological and biophysical dynamics is essential for understanding organoid physiology and its perturbation under external stimuli [16]–[19] .

Most existing approaches for organoid analysis rely on confocal fluorescence microscopy [20]. While fluorescence imaging offers molecular specificity and high spatial resolution, it typically requires exogenous labeling, fixation, or genetic modification, which limits clinical translatability and complicates long-term live imaging. Moreover, repeated volumetric fluorescence imaging introduces phototoxicity and photobleaching, constraining the observation of slow or prolonged dynamic processes [21], [22]. These limitations motivate the development of label-free imaging modalities that enable long-term, quantitative monitoring of live organoids [23]–[25].

Holotomography (HT) is a label-free 3D quantitative phase imaging technique that reconstructs volumetric refractive-index (RI) distributions of transparent specimens [26], [27]. Because RI is directly related to local protein concentration and cellular dry mass, HT provides intrinsic biophysical contrast without the need for labels [28]–[31], making it particularly attractive for longitudinal imaging of living organoids. Recent studies have demonstrated the feasibility of long-term HT imaging of organoids with minimal phototoxicity [32]–[34]. However, the broader adoption of HT for quantitative 3D analysis remains limited by fundamental optical constraints.

A major challenge in HT arises from incomplete spatial-frequency coverage due to finite illumination and detection numerical apertures. This limitation, commonly referred to as the missing-cone problem, leads to anisotropic resolution and axial distortion in reconstructed RI tomograms [35]. Missing-cone artifacts primarily affect low-spatial-frequency components and compromise accurate delineation of epithelial layers, lumen geometry, and RI-based quantitative measurements—features that are central to organoid analysis. Although various reconstruction and regularization strategies have been proposed to mitigate missing-cone artifacts [35]–[37],

This work was supported by National Research Foundation of Korea grant funded by the Korea government (MSIT) (RS-2024-00442348, 2022M3H4A1A02074314), Korea Institute for Advancement of Technology (KIAT) through the International Cooperative R&D program (P0028463), the Korean Fund for Regenerative Medicine (KFRM) grant funded by the Korea government (the Ministry of Science and ICT and the Ministry of Health & Welfare) (21A0101L1-12), and the Samsung Research Funding Center of Samsung Electronics under Grant (SRFC-IT1401-08).
*(Corresponding author: YongKeun Park.)*

ChulMin Oh and Juyeon Park are with the Department of Physics and the KAIST Institute for Health Science and Technology, Korea Advanced Institute of Science and Technology (KAIST), Daejeon, Republic of Korea (e-mail: chisquare@kaist.ac.kr; jooyunpark95@kaist.ac.kr).

Jimin Cho is with the Graduate School of Stem Cell and Regenerative Biology and the KAIST Institute for Health Science and Technology, KAIST, Daejeon, Republic of Korea (e-mail: a01077271401@kaist.ac.kr).

Hoyeon Lee is with Tomocube Inc., Daejeon, Republic of Korea (e-mail: hoyeon.lee@tomocube.com).

YongKeun Park is with the Department of Physics, the KAIST Institute for Health Science and Technology, and the Graduate School of Stem Cell and Regenerative Biology, KAIST, and also with Tomocube Inc., Daejeon, Republic of Korea (e-mail: yk.park@kaist.ac.kr).

these approaches often prioritize visual fidelity or resolution enhancement and may not directly address the robustness of downstream quantitative analysis. More rigorous reconstruction frameworks based on nonlinear inverse-scattering models, such as Lippmann–Schwinger formulations, have been shown to substantially improve reconstruction accuracy in optical diffraction tomography [38], but their computational complexity and practical constraints limit their applicability to large-volume, long-term live organoid imaging.

Recent advances in medical imaging have explored data-driven and physics-based strategies to recover missing information in tomographic systems [39], including deep-learning–based regularization and angular recovery methods [40], [41]. While these approaches can improve apparent isotropy, they typically require extensive training data, high computational cost, or strong model assumptions, which may limit their applicability to large-volume, long-term organoid imaging under realistic experimental conditions. Consequently, there remains a need for analysis frameworks that explicitly acknowledge the physical limitations of HT while enabling robust, quantitative characterization of organoid morphology and biophysical properties.

In this work, we propose a quantitative analysis framework that addresses the missing-cone problem at the level of image representation rather than reconstruction. By selectively emphasizing spatial-frequency components that are less sensitive to missing-cone–induced distortion, we introduce morphology-preserving holotomography (MP-HT), a torus-shaped spatial filtering strategy that stabilizes 3D morphological features in RI tomograms. Building on this representation, we develop a multi-stage 3D segmentation pipeline tailored to organoid geometry and a model-based approach for quantifying dry-mass density and total dry mass that explicitly incorporates the imaging point spread function.

Using this framework, we perform long-term, label-free imaging of live hepatic organoids undergoing expansion, collapse, and fusion. We demonstrate that the proposed approach enables consistent segmentation of epithelial and luminal structures, robust morphometric analysis across diverse geometries, and morphology-independent quantification of dry mass. Together, these results establish a practical pathway for extending HT from qualitative visualization toward quantitative, longitudinal analysis of complex 3D multicellular systems.

## II. Methods

### A. Morphology enhancement in HT via torus-shaped spatial filtering

HT reconstructs the 3D RI distribution of a transmissive specimen by solving an inverse light-propagation problem under systematically modulated illumination [42], [43]. As incident light traverses the specimen, it is refracted according to the local RI variations, and the resulting intensity measurements are computationally inverted to obtain a volumetric RI tomogram. In the HT implementation used in this study (Fig. 1a), the specimen is illuminated by spatially modulated incoherent light from an LED source, while the objective lens is axially scanned to acquire volumetric intensity stacks. Multiple illumination patterns are sequentially projected onto the pupil plane to encode angular diversity and enable RI reconstruction [27], [44], [45].

Due to the finite numerical apertures (NAs) of the illumination and detection optics, the spatial-frequency components accessible to HT are inherently limited. In particular, the spatial-frequency support of HT lacks a conical region along the axial frequency direction, commonly referred to as the missing cone (Fig. 1b). This anisotropic deficiency primarily affects low-spatial-frequency components and gives rise to missing-cone artifacts, which manifest as axial elongation and distortion of structures in reconstructed RI tomograms [35], [46]. Such distortions compromise both quantitative morphology and RI-based measurements, especially in thick 3D samples such as organoids.

Rather than attempting to reconstruct the missing spatial-frequency information, we address this limitation at the level of image representation by selectively emphasizing spatial-frequency components that are less sensitive to missing-cone–induced distortions. Specifically, we apply a torus-shaped band-pass filter in the 3D Fourier domain of the RI tomogram to suppress low-frequency components while preserving high-spatial-frequency RI texture that encodes local morphological features.

The torus-shaped filter is defined as

$$f\left(k_x,k_y,k_z\right)=\begin{cases} g\left(1-\dfrac{\left(k_{x,y}-k_0\right)^2}{\beta^2}-\dfrac{k_z^2}{\alpha^2}\right), & g\left(x\right)>0, \\ 0, & \textit{otherwise}, \end{cases} \tag{1}$$

where $k_x$, $k_y$, $k_z$ denote the spatial frequecies along the $x$, $y$, and $z$ directions, respectively, and $k_{x,y}=\sqrt{k_x^2+k_y^2}$. The geometric parameters $\alpha$ and $\beta$ defines the lateral and axial extents of the torus and are derived from the theoretical spatial-frequency support of the HT system, which is determined by the illumination and detection NAs and the illumination wavelength (Fig. 1c). The detailed derivation of these parameters is provided in Appendix A.

Following application of the torus filter, the filtered RI volume is squared and smoothed using a 2D Gaussian kernel with a standard deviation of 1/(8$\alpha$ × lateral pixel size)). This nonlinear transformation enhances morphology-related high-frequency features and improves robustness for subsequent segmentation. The resulting representation, referred to as MP-HT, no longer exhibits the missing-cone structure in the Fourier domain that is present in the original RI tomogram (Fig. 1b).

The physical basis of MP-HT can be understood by considering the distinct responses of different RI components to missing-cone filtering. Low-spatial-frequency RI offsets, which encode bulk density variations over large length scales, are strongly distorted by missing-cone limitations and give rise to axial elongation. In contrast, high-spatial-frequency RI texture—arising from heterogeneous subcellular structures distributed throughout the cellular volume—remains largely unaffected by the missing cone and preserves the true morphology of the specimen [47]. By preferentially retaining these texture-dominated components, MP-HT provides a morphology-stable representation that is well suited for quantitative 3D analysis.

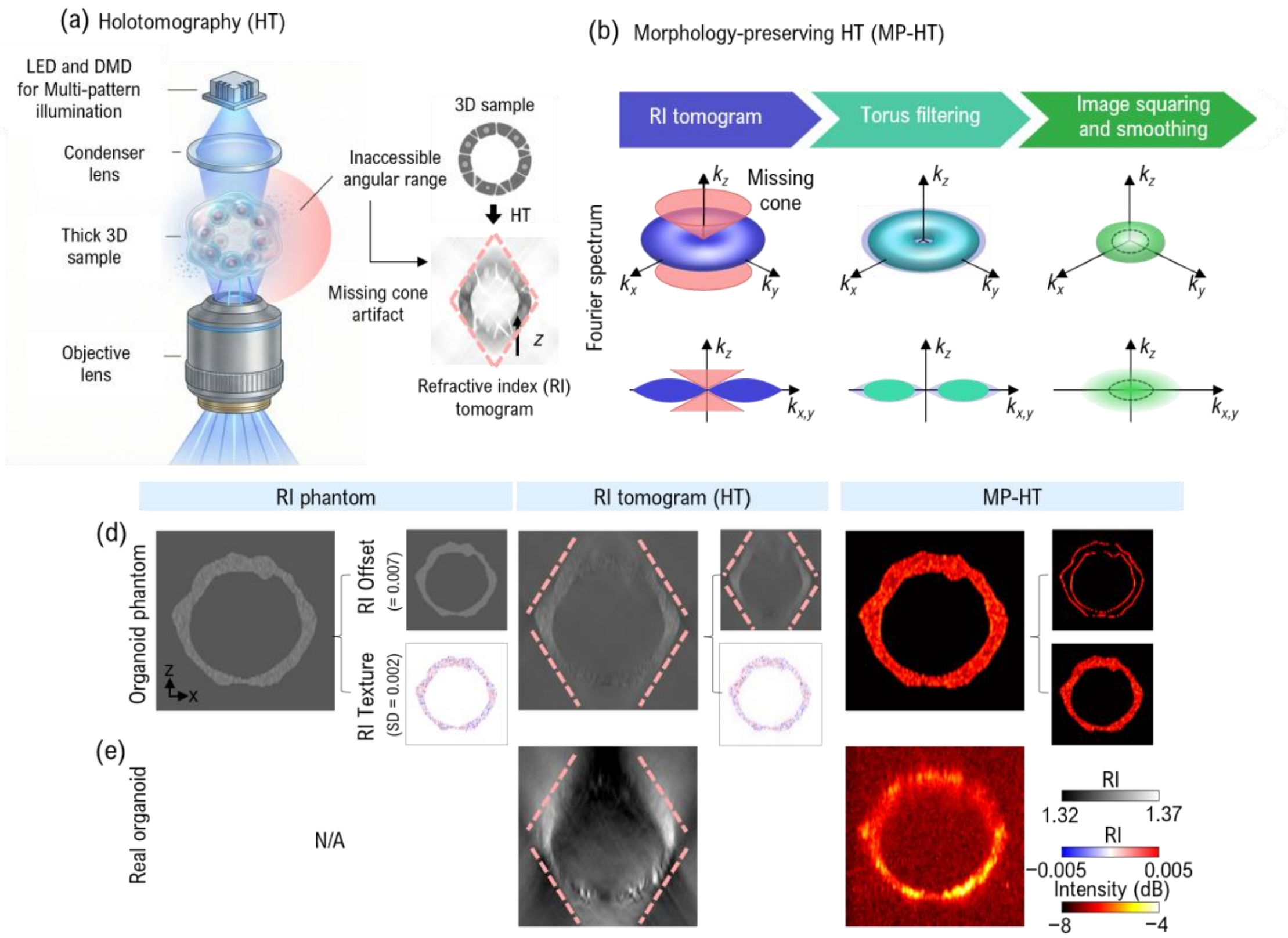

**Fig. 1. HT principle, missing-cone artifact, and motivation for morphology-preserving analysis.** (a) Schematic of the HT system based on multi-pattern illumination. The finite angular ranges of illumination and detection optics lead to incomplete sampling of the three-dimensional spatial-frequency domain. (b) This incomplete coverage results in a missing-cone region in Fourier space (red), which primarily affects low-spatial-frequency components and causes axial elongation and distortion in reconstructed RI tomograms. (c) To address this limitation at the level of image representation, a torus-shaped band-pass filter is defined from the spatial-frequency support of the HT system (blue) and applied to the RI tomogram. The filtered image is subsequently squared and Gaussian-smoothed to enhance morphology-related high-frequency features; the dashed contour indicates the half-maximum surface of the resulting spectrum. (d) Simulation results using a synthetic organoid phantom show that uniform RI offsets are strongly distorted by the missing-cone artifact, whereas high-spatial-frequency RI texture preserves the underlying morphology. (e) Experimental organoid data demonstrate the same trend, supporting the use of MP-HT as a robust representation for quantitative morphological analysis under missing-cone–limited conditions. Regions affected by missing-cone–induced distortion are indicated by red dashed outlines.

To illustrate this principle, we generated a synthetic organoid phantom composed of two RI components confined to the same geometry: (i) a uniform RI offset ($\Delta n$ = 0.007) representing elevated protein density, and (ii) a Gaussian random RI texture with a standard deviation of 0.002 emulating subcellular heterogeneity (Fig. 1d). When the spatial-frequency support of HT was applied to simulate an RI tomogram, the uniform-offset component exhibited pronounced axial elongation due to the missing cone, whereas the texture component preserved the original geometry. Applying the proposed torus-shaped filtering selectively emphasized the texture component, thereby mitigating axial distortion and stabilizing the apparent morphology. The same behavior was observed in experimental organoid measurements (Fig. 1e), demonstrating that MP-HT consistently emphasizes spatial-frequency components that are robust to missing-cone artifacts.

Finally, we validated the biological relevance of MP-HT by comparing it with confocal fluorescence images of plasma membranes labeled with CellMask in hepatic and intestinal organoids (Fig. 2). Across both organoid types, MP-HT closely recapitulated epithelial boundaries observed in fluorescence images, particularly in regions affected by axial distortion in RI tomograms. Together, these results establish MP-HT as a representation optimized for morphology-preserving analysis of HT data under missing-cone–limited conditions, forming the basis for the segmentation and quantitative analysis described in the following sections.

### *B. Segmentation pipeline*

#### *1) Whole-organoid segmentation*

The first step of the segmentation pipeline aims to obtain a robust 3D binary mask representing the entire organoid volume. This task is particularly challenging in HT due to missing-cone–induced axial distortion, reduced contrast at depth, and intensity spreading introduced by frequency-domain filtering. Leveraging the observation that MP-HT selectively enhances organoid boundaries while suppressing low-frequency artifacts (Fig. 3a,b), we designed a whole-organoid segmentation pipeline tailored to MP-HT images (Fig. 3e).

The pipeline consists of four sequential stages: (i) morphology-preserving transformation, (ii) image preprocessing, (iii) initial binary mask generation, and (iv) boundary-thickness correction via phantom-based erosion.

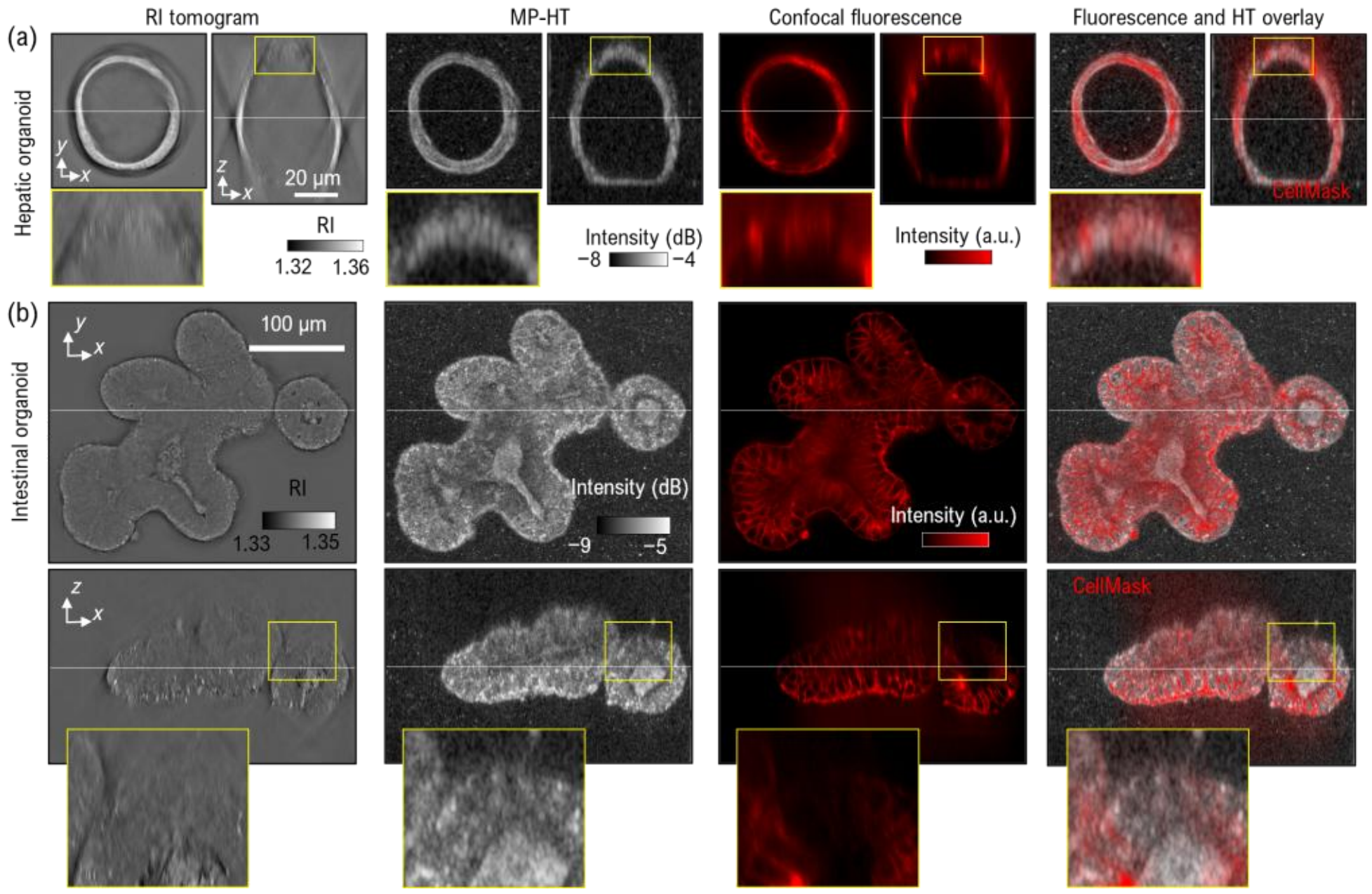

**Fig. 2. Comparison of MP-HT with confocal fluorescence imaging.** Representative results are shown for (a) hepatic organoids and (b) intestinal organoids. For each case, reconstructed R) tomograms (left), corresponding MP-HT images (middle), confocal fluorescence images acquired using a plasma membrane stain (CellMask; red), and their overlays (right) are displayed.

**(i) Morphology-Preserving Transformation:** The reconstructed RI tomogram is first transformed into an MP-HT representation using the torus-shaped spatial filtering procedure described in Section II-A. This step enhances high-spatial-frequency features associated with organoid boundaries while suppressing low-frequency components that contribute to axial elongation. As a result, the organoid contour becomes more pronounced and spatially consistent across axial slices, providing a suitable input for subsequent segmentation steps.

**(ii) Image Preprocessing:** To facilitate robust volumetric segmentation, the MP-HT image undergoes several preprocessing operations. First, the intensity values are converted to a logarithmic scale to compress dynamic range and stabilize thresholding. The volume is then downsampled such that the lateral pixel spacing matches the axial spacing, yielding approximately isotropic voxels. A 3D median filter with a kernel size of 5×5×5 voxels is applied to suppress speckle-like noise while preserving boundary features. After filtering, the volume is upsampled back to the original resolution.

To mitigate stitching artifacts and background offsets along the axial direction, we subtract, at each lateral pixel location, the minimum intensity value along the axial axis. This operation effectively normalizes axial intensity variations without affecting boundary-enhanced features.

**(iii) Initial Binary Mask Generation:** An initial whole-organoid mask is generated by applying Otsu's global thresholding method [48] to the preprocessed MP-HT image. Otsu thresholding provides an objective and parameter-free means of separating boundary-enhanced organoid regions from the background. The resulting binary mask reliably captures the overall organoid shape but tends to overestimate the true biological boundary due to resolution loss and intensity spreading introduced by filtering and thresholding.

**(iv) Boundary-Thickness Correction via Phantom-Based Erosion:** To compensate for the artificially thick boundary layer present in the initial mask, we employ a phantom-based boundary-thickness correction strategy. Specifically, a synthetic RI phantom is constructed to reproduce the statistical properties of the segmented organoid. Using the initial mask, we compute (i) the mean RI offset and (ii) the standard deviation of local RI texture, where the latter is evaluated within a 5×5×1 voxel window and averaged over the masked region. Assuming a normal distribution, these statistics are combined with the initial mask to synthesize a 3D phantom RI volume (Fig. 3e).

The phantom volume is then processed using the same steps (i)–(iii) described above, yielding a corresponding phantom mask. The excess boundary thickness is quantified by identifying the 3D erosion operator—parameterized by an ellipsoidal structuring element—that minimizes the discrepancy between the eroded phantom mask and the original mask used to generate the phantom. The discrepancy is evaluated using an intersection-over-union metric, and the detailed procedure for selecting the optimal structuring element is provided in Appendix B.

Finally, the same erosion operator is applied to the initial organoid mask, resulting in a corrected whole-organoid segmentation that more accurately reflects the underlying biological boundary. This final 3D binary mask is used for quantitative morphometric analysis and serves as the input for the subsequent lumen–epithelium separation step.

### 2) *Lumen-epithelium separation*

In many organoid systems, internal luminal spaces are enclosed by a continuous epithelial layer, making accurate separation of the lumen from the epithelium essential for physiologically meaningful morphometric analysis [49], [50]. In the ductal-like hepatic organoids analyzed in this study, cystic lumina are surrounded by epithelial cells, resulting in relatively low signal intensity within the lumen region in MP-HT images (Fig. 2a). Under ideal conditions, whole-organoid segmentation would therefore yield an epithelium mask that fully encloses the lumen, allowing the lumen to be inferred as the interior region.

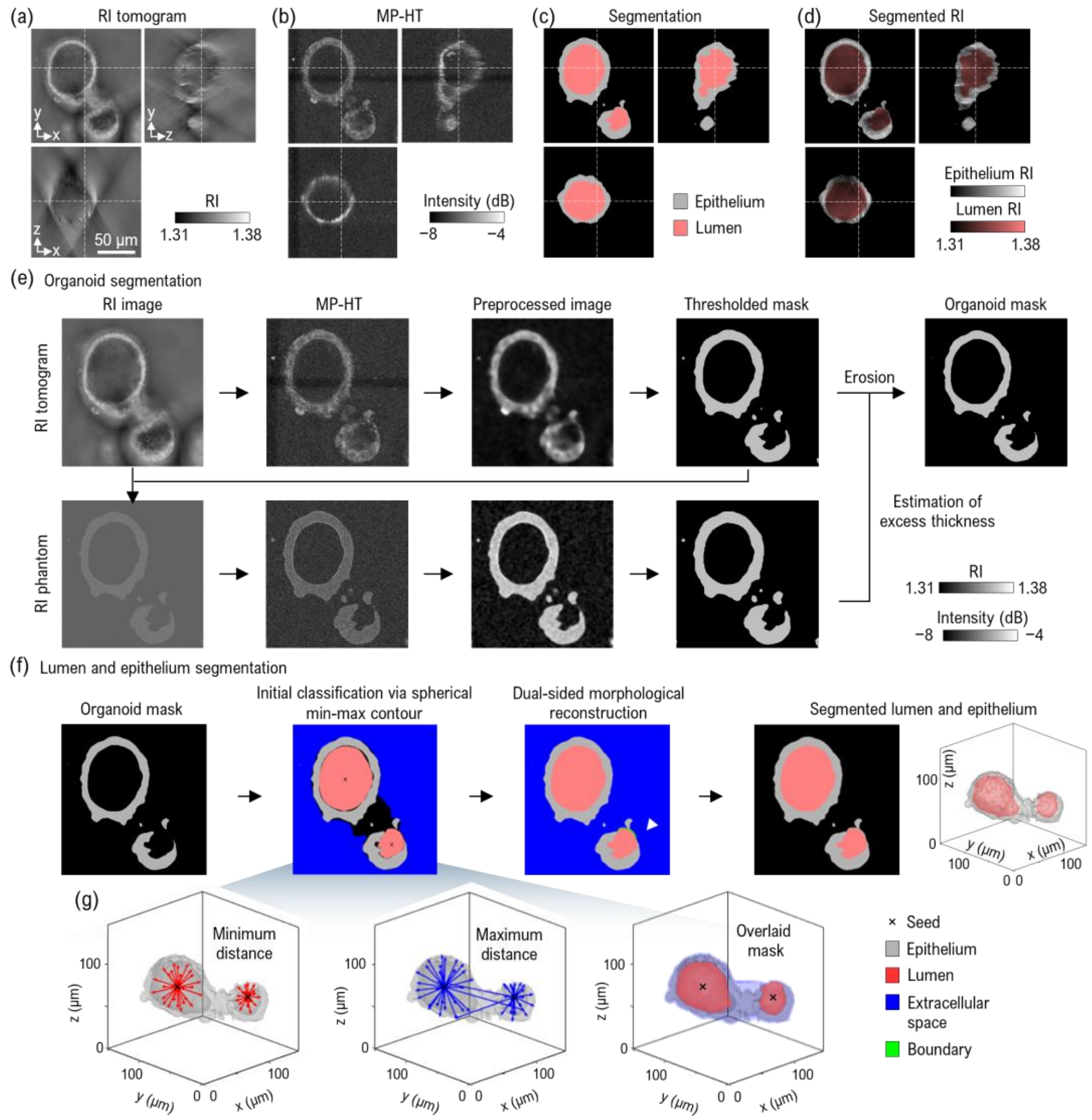

**Fig.3. Simulation study for RI and dry-mass quantification under missing-cone–limited conditions.** (a) Ellipsoidal shell phantoms with varying axial aspect ratios (0.5–1.5) were generated to model organoid-like morphologies. (b) Corresponding RI tomograms were simulated by convolving each phantom with the PSF of the HT system. (c) The proposed MP-HT–based segmentation pipeline was applied to obtain volumetric masks of the phantoms. (d) Segmentation masks were subsequently convolved with the same PSF to construct the forward model used for RI estimation. (e) Quantitative results for mean RI and total dry mass as a function of the aspect ratio. Ground-truth values are indicated by dashed lines. Results obtained using the conventional voxel-averaging method within the segmentation mask are shown in red, whereas results obtained using the proposed model-based approach incorporating PSF-convolved masks are shown in blue.

In practice, however, multiple scattering, depth-dependent contrast loss, and residual missing-cone effects introduce holes and discontinuities in the epithelial boundary. These artifacts prevent reliable lumen identification using conventional approaches such as morphological filling or connected-component analysis, which implicitly assume a closed epithelial surface. To robustly separate the lumen and epithelium under such conditions, we developed a multi-stage segmentation pipeline that explicitly reconstructs missing epithelial boundaries using geometric constraints (Fig. 3f).

The proposed pipeline consists of four stages: (i) lumen seed detection, (ii) spherical min–max contour estimation for initial voxel classification, (iii) dual-sided morphological reconstruction, and (iv) final mask assembly.

**(i) Lumen Seed Detection:** Candidate lumen centers are identified using a 3D distance transform computed from the epithelium mask obtained in the previous step. Local maxima of the distance map correspond to regions that are maximally distant from the epithelial boundary and thus likely to lie within luminal cavities. Extended-maxima detection is applied to extract such candidate regions. Components touching the image border are removed to avoid spurious detections, and only components fully contained within the bounding box of the organoid mask are retained. The remaining candidates are sorted by volume, and the centroids of up to the three largest components are selected as lumen seeds.

**(ii) Spherical Min–Max Contour Estimation:** For each lumen seed, we perform spherical ray sampling to estimate the inner and outer extents of the epithelial boundary (Fig. 3g).

Rays are propagated from the seed point across a dense set of azimuth–elevation directions, accounting for anisotropic voxel dimensions via an axial scaling factor. Along each ray, we measure the minimum and maximum distances at which the ray intersects the epithelium mask, yielding two angularly parameterized distance functions: a minimum-radius surface approximating the inner epithelial boundary and a maximum-radius surface approximating the outer boundary.

Missing angular samples caused by mask discontinuities are filled by linear interpolation followed by median filtering to obtain smooth, continuous surfaces. Based on these surfaces, voxels are initially classified as lumen candidates if their radial distance is smaller than the minimum-radius surface, and as extracellular candidates if their distance exceeds the maximum-radius surface. Candidate classifications from all seeds are aggregated to form initial lumen and extracellular masks, excluding voxels belonging to the original organoid mask.

To bridge gaps in the epithelial boundary separating the lumen and extracellular regions, we perform constrained dual-sided morphological reconstruction. Starting from the initial lumen and extracellular masks, both regions are iteratively dilated using a 3×3×3 structuring element, while preventing expansion into each other or into the original organoid mask. Voxels at which the advancing fronts of the two regions meet are labeled as boundary voxels. The reconstruction proceeds until the lumen region stabilizes, effectively repairing missing epithelial walls and restoring a closed epithelial surface.

This dual-sided formulation ensures that reconstructed boundaries are consistent with both the lumen and extracellular regions, avoiding the bias that would result from unilateral filling operations.

**(iv) Final Mask Assembly:** The final epithelium mask is obtained by combining the original epithelium mask with the reconstructed boundary voxels, while the stabilized lumen region defines the lumen mask. Together, these masks yield a complete and topologically consistent separation of lumen and epithelium throughout the 3D organoid volume.

Following automated segmentation, detected lumens are manually reviewed to remove biologically irrelevant structures, such as spurious cavities arising from noise or segmentation artifacts. The final segmentation results are shown in Fig. 3c,d alongside the original RI tomogram and MP-HT image. Full implementation details, including parameter settings, are provided in Appendix C.

### *C. Model-based quantification of dry mass and dry-mass density*

RI tomograms obtained by HT provide direct access to the optical density of biological specimens and thus enable estimation of cellular dry mass. However, quantitative RI analysis is fundamentally challenged by missing-cone–induced distortions, which affect both the apparent morphology and the voxel-wise RI values. As a consequence, the conventional approach of computing the mean RI within a segmented volume yields morphology-dependent and inconsistent estimates when applied to HT data, particularly for objects with varying axial aspect ratios [51].

To address this limitation, we adopt a model-based quantification strategy that explicitly incorporates the image-formation characteristics of HT through its PSF. Rather than relying on voxel-wise RI values, this approach estimates intrinsic RI parameters by fitting a forward model that accounts for spatial-frequency limitations, thereby decoupling RI quantification from morphology-dependent distortions.

**1) Forward Model of RI Tomography:** The effective PSF of HT is defined as the inverse Fourier transform of the system's spatial-frequency support, which is determined by the illumination and detection numerical apertures and the illumination wavelength (Appendix A). Under the single-scattering approximation, the measured RI tomogram can be modeled as the convolution of the intrinsic RI distribution with this PSF.

We approximate the intrinsic RI distribution of an organoid as the sum of a uniform background RI and a uniform RI offset confined to the epithelial region:

$$n(x,y,z) = n_{\mathrm{bg}} + \Delta n_{\mathrm{org}} \cdot M(x,y,z), \tag{2}$$

where $n_{bg}$ denotes the extracellular RI, $\Delta n_{org}$ is the RI offset of the organoid relative to the background, and $M(x,y,z)$ is a binary epithelium mask obtained from the segmentation pipeline.

Convolving both sides of Eq. (2) with the PSF yields

$$n_{HT}(x,y,z) = n_{\mathrm{bg}} + \Delta n_{\mathrm{org}} \cdot \left[ M(x,y,z) * \mathrm{PSF}(x,y,z) \right], \tag{3}$$

where $n_{HT}(x,y,z)$ represents the measured HT tomogram. In this formulation, the convolved mask term is fully determined by segmentation and system optics, leaving $n_{bg}$ and $\Delta n_{org}$ as the only unknown scalar parameters.

**2) Parameter Estimation and Dry-Mass Calculation:** The parameters $n_{bg}$ and $\Delta n_{org}$ are estimated by solving Eq. (3) using a least-squares approach over all voxels in the volume. This global fitting procedure effectively averages out local RI fluctuations and reduces sensitivity to morphology-dependent distortions introduced by the missing cone.

The dry-mass density $\rho_{dm}$ is then computed from the estimated RI offset using the established linear relationship

$$\rho_{dm} = \Delta n_{\mathrm{org}} / RII, \tag{4}$$

where $RII$ is a refractive index increment, the proportionality constant based on classical refractometry of biological materials ($RII$ = 0.185 mL/g) [52], [53]. The total dry mass is obtained by multiplying the dry-mass density by the segmented epithelial volume.

This model-based strategy yields dry-mass estimates that are inherently independent of organoid shape, provided that the segmentation mask accurately captures the epithelial region.

**3) Validation Using Synthetic Phantoms:** To evaluate the robustness of the proposed quantification method, we performed simulations using synthetic organoid phantoms with varying axial aspect ratios (Fig. 4a). Each phantom consisted of an ellipsoidal shell geometry with a known RI offset, allowing direct comparison with ground-truth values.

Simulated RI tomograms were generated by convolving the intrinsic RI distributions with the same PSF used in the forward model (Fig. 4b). The phantoms were then segmented using the proposed MP-HT–based pipeline (Fig. 4c), and the corresponding masks were convolved with the PSF to construct the model term in Eq. (3) (Fig. 4d).

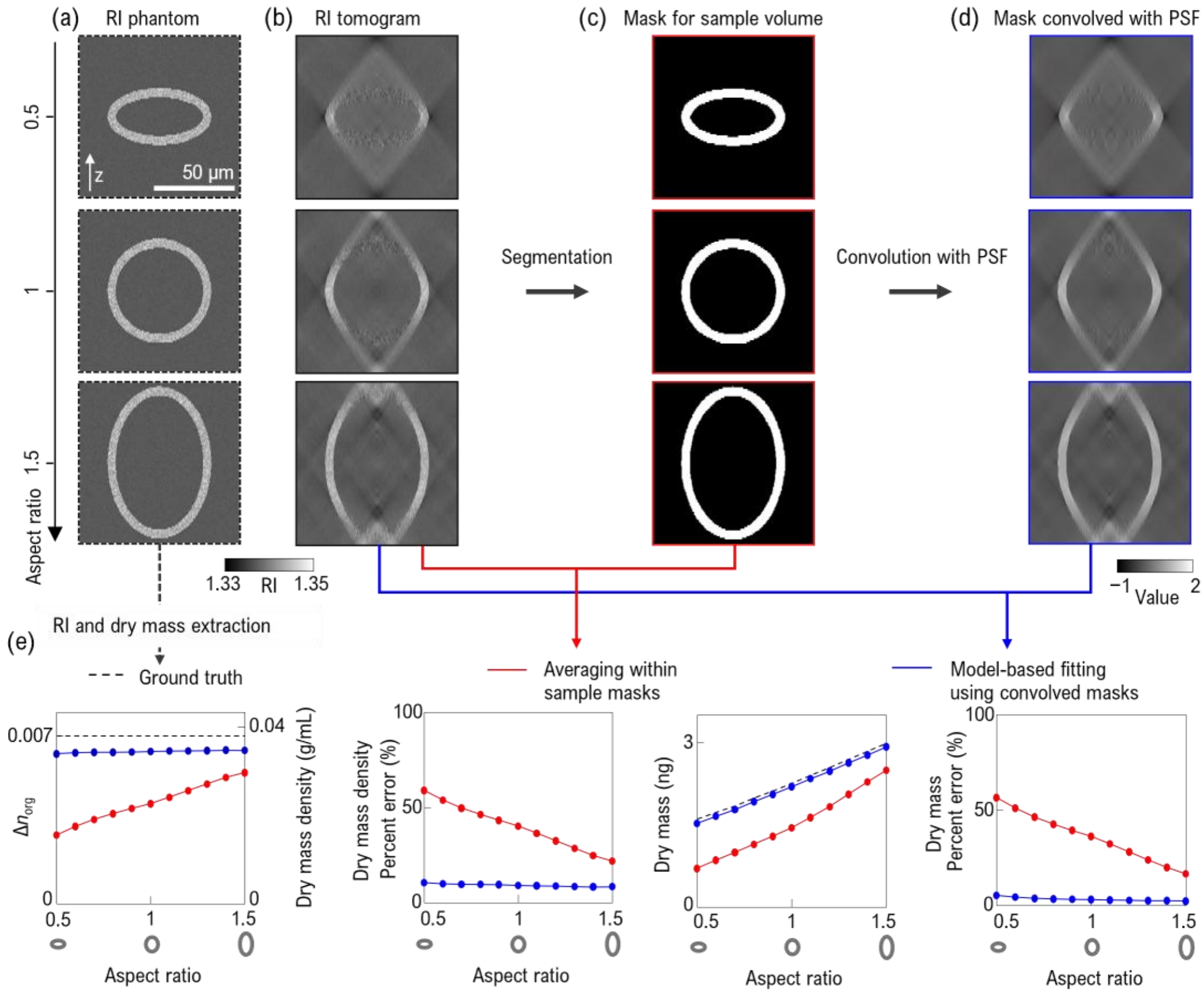

**Fig. 4. Simulation study for RI and dry-mass quantification under missing-cone–limited conditions.** (a) Ellipsoidal shell phantoms with varying axial aspect ratios (0.5–1.5) were generated to model organoid-like morphologies. (b) Corresponding RI tomograms were simulated by convolving each phantom with the PSF of the HT system. (c) The proposed MP-HT–based segmentation pipeline was applied to obtain volumetric masks of the phantoms. (d) Segmentation masks were subsequently convolved with the same PSF to construct the forward model used for RI estimation. (e) Quantitative results for mean RI and total dry mass as a function of the aspect ratio. Ground-truth values are indicated by dashed lines. Results obtained using the conventional voxel-averaging method within the segmentation mask are shown in red, whereas results obtained using the proposed model-based approach incorporating PSF-convolved masks are shown in blue.

As shown in Fig. 4e, the conventional voxel-averaging method exhibited strong morphology-dependent deviations in both RI and dry-mass estimates as the aspect ratio varied. In contrast, the proposed model-based approach produced stable RI and dry-mass values across all tested geometries, achieving less than 10% error in total dry-mass estimation. A slight underestimation of RI was observed, which can be attributed to modest over-segmentation of the epithelial mask (approximately 7% volumetric excess). Importantly, this bias remained consistent across morphologies, preserving the relative accuracy of temporal measurements.

**4) Scope and Limitations:** The uniform RI-offset assumption employed in Eq. (2) represents a first-order approximation of the epithelial region and is appropriate for organoid-scale analysis, where the primary interest lies in bulk biophysical properties rather than subcellular heterogeneity. Local RI variations arising from intracellular organelles or compositional gradients are effectively averaged out by the global fitting procedure and are not explicitly modeled.

Furthermore, the present approach assumes the validity of the single-scattering approximation underlying HT reconstruction. In thicker samples, multiple scattering may lead to reduced RI contrast and potential underestimation of dry mass, an effect that is partially mitigated—but not eliminated—by the proposed model. These limitations are discussed in more detail in Section IV.

### *D. Sample preperation*

Mouse hepatic organoids (mHOs) were purchased from STEMCELL Technologies (Vancouver, Canada), embedded in Matrigel domes, and cultured in HepatiCult™ Organoid Growth Medium (Mouse; STEMCELL Technologies) according to the manufacturer's protocol. Human gastrointestinal organoids (3dGRO; Merck KGaA, cat. no. SCC734) were similarly embedded in Matrigel domes and cultured in IntestiCult™ medium (STEMCELL Technologies) following the manufacturer's instructions.

Free fatty acids (FFAs)—oleic acid, linoleic acid, and palmitic acid—were dissolved in 100% ethanol to prepare 100 mM stock solutions. A 10% (w/v) bovine serum albumin (BSA) solution in phosphate-buffered saline (PBS) was mixed with each FFA stock solution at a 1:9 (v/v) ratio to generate FFA–BSA conjugates. These conjugates were diluted in organoid culture medium to final concentrations of 200 μM (linoleic acid) and 400 μM (palmitic acid and oleic acid). Under these conditions, treatment with linoleic acid, palmitic acid, and oleic acid reproducibly induced organoid expansion, collapse, and fusion dynamics, respectively.

### E. Holotomography Imaging

Holotomographic imaging was performed using HT-X1™ and a customized X1 system (Tomocube Inc., Daejeon, Republic of Korea). The HT-X1 system was equipped with a condenser lens of NA 0.72 and a 0.95 NA air objective, whereas the customized system employed a 1.1 NA water-immersion objective. Illumination wavelengths were set to 460 nm for the HT-X1 system and 660 nm for the customized system.

Time-lapse HT imaging was conducted in a temperature- and $CO_2$-controlled incubation chamber maintained at 37 °C and 5% $CO_2$. Volumetric RI tomograms were acquired at 6-h intervals for up to 30 h. For each time point, multiple illumination patterns were sequentially projected to enable RI reconstruction, as described in Section II-A. All HT data analyzed in this study were acquired under identical imaging parameters for a given experimental condition.

### F. Confocal Fluorescence Imaging for Validation

For fluorescence-based validation of epithelial morphology, both hepatic and gastrointestinal organoids were incubated with CellMask™ Deep Red plasma membrane stain (Thermo Fisher Scientific, Waltham, MA, USA) at a final concentration of 5 μg/mL in culture medium for 1 h at 37°C under 5% $CO_2$. Following incubation, confocal fluorescence imaging was performed using a spinning-disk confocal microscope (X-Light V2; CrestOptics, Rome, Italy) with 640 nm excitation and a 700 nm emission filter.

Confocal images were acquired to qualitatively assess epithelial boundary correspondence and were not used for quantitative analysis. All fluorescence measurements were performed on separate samples from those used for long-term HT imaging to avoid phototoxic effects.

### G. Code and Data Availability

The source code supporting this study, including implementations of MP-HT, the 3D segmentation pipeline, and the model-based RI quantification, together with the organoid datasets used in Fig. 3, is publicly available at the following GitHub repository:: *https://github.com/BMOLKAIST*.

## III. Results

To demonstrate the utility of the proposed quantitative analysis framework, we analyzed live hepatic organoids undergoing three representative dynamic processes: expansion, collapse, and fusion. In all cases, organoids were segmented using the proposed MP-HT–based 3D pipeline, and RI–derived biophysical parameters were quantified using the model-based approach described in Section II-C. The extracted features include epithelial and lumen volumes, surface areas, sphericity, lumen aspect ratio, dry-mass density, and total dry mass.

### A. Expanding Organoid

We first investigated the expansion dynamics of hepatic organoids over a 30-h time course. Representative results are shown in Fig. 5, including 3D segmentations of the epithelium and lumen, corresponding geometric descriptors, and biophysical measurements.

Time-lapse analysis revealed continuous increases in both epithelial and lumen volumes (Fig. 5c), while the overall epithelium–lumen architecture remained approximately concentric and spherical throughout the experiment (Fig. 5a,b). Initially, the lumen occupied a small fraction of the organoid volume; however, after approximately 6 h, the lumen-to-organoid volume fraction stabilized at around 0.5, indicating coordinated growth of the lumen and epithelium (Fig. 5e). A similar trend was observed for surface areas: both apical (lumen-facing) and basal (outer) surfaces increased proportionally, leading to a stable apical-to-basal surface-area ratio over time (Fig. 5d,e).

Line-profile analysis showed that the lumen aspect ratio progressively approached unity (Fig. 5f), indicating that the lumen became increasingly spherical despite gravitational loading. This behavior suggests mechanically stable expansion under the applied culture conditions.

In addition to morphological changes, biophysical analysis revealed a monotonic increase in total dry mass, reaching approximately a 4.5-fold increase over 30 h (Fig. 5g), consistent with accumulation of intracellular protein content during mitotic proliferation. In contrast, the dry-mass density remained nearly constant throughout the experiment, indicating preservation of cellular density homeostasis during coordinated volumetric growth.

### B. Collapsing Organoid

Next, we examined organoid collapse induced by a lipotoxic agent, with representative results shown in Fig. 6. In contrast to the expanding case, the lumen volume decreased rapidly and disappeared at approximately 18 h (Fig. 6a,b).

While the epithelial volume initially continued to increase, its growth stagnated several hours after lumen collapse (Fig. 6c). Apical and basal surface areas followed trends similar to those of the epithelial and lumen volumes (Fig. 6d), reflecting a coupled remodeling of surface geometry and volume.

As the lumen diminished, the organoid progressively deviated from a spherical morphology. Organoid sphericity, defined as $\pi^{1/3}\left(6V\right)^{2/3}/A$, decreased from approximately 0.85 to 0.5, whereas the lumen sphericity remained relatively preserved until collapse (Fig. 6e). The lumen volume fraction and the apical-to-basal surface-area ratio—both of which were comparable to stabilized values observed in the expanding organoid—decreased to zero following lumen collapse (Fig. 6f), indicating a breakdown of lumen–epithelium morphological homeostasis under cytotoxic stress.

Biophysical measurements further supported pathological progression. Dry-mass density decreased markedly during collapse (Fig. 6g), consistent with cytoplasmic swelling during cell death [41], [42]. Total dry mass also decreased, followed by a modest rebound after lumen collapse, suggesting complex intracellular remodeling rather than simple mass loss.

### C. Fusing Organoid

To evaluate performance in more complex geometries, we analyzed fusing hepatic organoids containing two distinct lumens. Representative results are shown in Fig. 7.

Segmentation results confirmed that the two lumens fused between 12 and 18 h (Fig. 7a,b). During fusion, both epithelial and lumen volumes exhibited transient increases followed by decreases (Fig. 7c), accompanied by corresponding changes in surface areas (Fig. 7d). The lumen volume fraction and apical-to-basal surface-area ratio showed short-lived increases during the fusion interval, indicating temporary epithelial thinning (Fig. 7e).

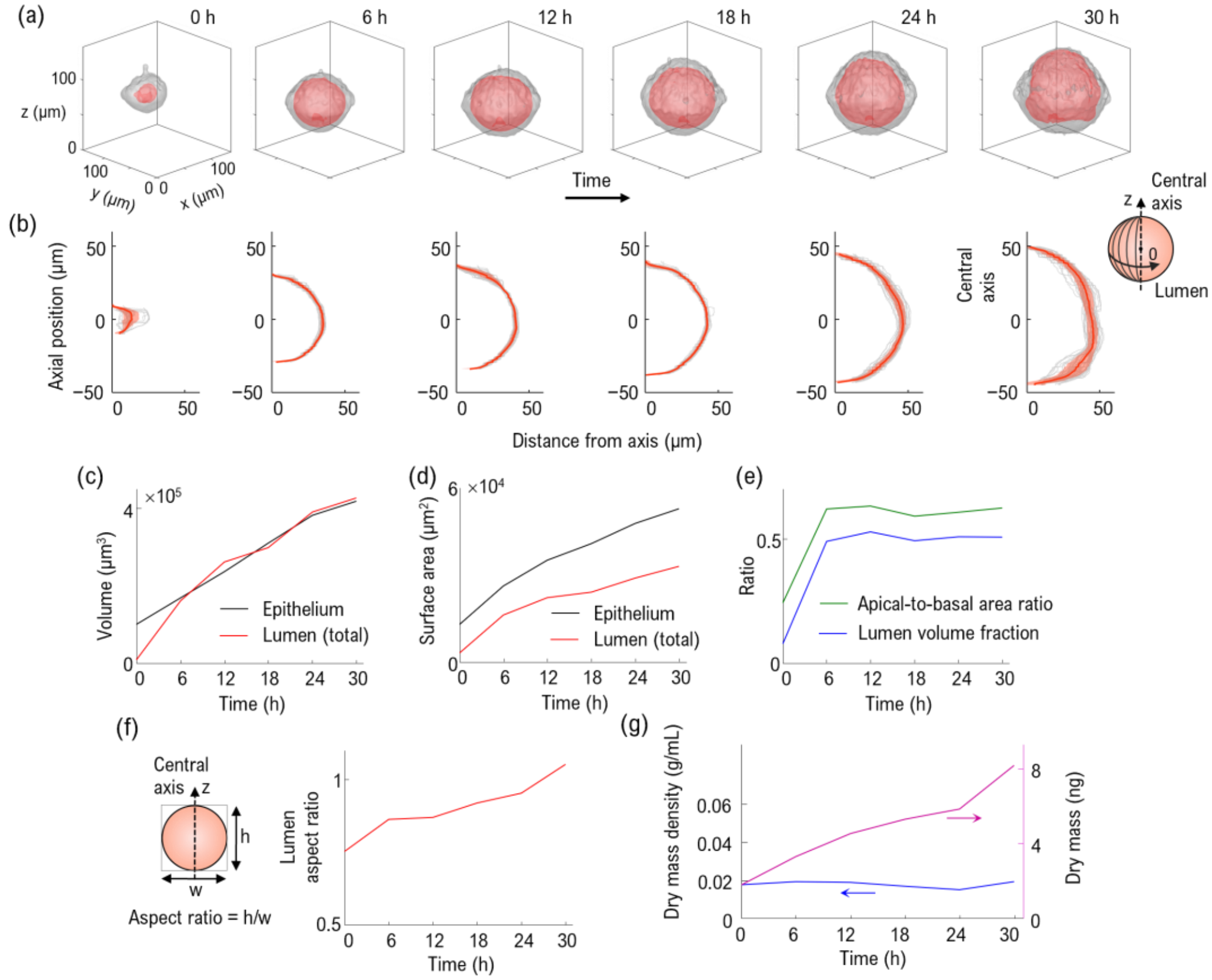

**Fig. 5. Quantitative analysis of an expanding organoid over a 30-h time course.** (a) Three-dimensional renderings of the segmented epithelium and lumen at representative time points. (b) Lumen line profiles extracted along the vertical axis passing through the lumen centroid. Gray curves correspond to individual radial profiles sampled from the lumen surface at 10° angular intervals, while the red curve denotes their mean and the shaded region indicates the standard deviation. Using the segmented epithelium and lumen masks, temporal changes in (c) epithelial and lumen volumes, (d) epithelial and lumen surface areas, and (e) the corresponding volume and surface-area ratios relative to the entire organoid were quantified. (f) The lumen aspect ratio was computed from the line-profile analysis shown in (b). (g) Dry-mass density (blue) and total dry mass (purple) estimated from the epithelial region are plotted as functions of time using dual y-axes.

The fusion process was asymmetric: the smaller lumen (lumen 2) was gradually absorbed into the larger lumen (lumen 1), as reflected by opposing volume trajectories (Fig. 7c). Aspect-ratio analysis corroborated this asymmetry, with lumen 1 progressively approaching a spherical geometry while lumen 2 became increasingly deformed prior to disappearance (Fig. 7f). After fusion, the merged lumen settled into an intermediate aspect ratio.

Dry-mass measurements provided additional insight into the biophysical changes accompanying fusion. Throughout the fusion process, total dry mass remained approximately constant, whereas dry-mass density decreased due to volumetric expansion (Fig. 7g). Following fusion, total dry mass showed a slight decrease while dry-mass density increased, suggesting redistribution of biomass driven primarily by morphological and volumetric dynamics rather than net mass accumulation or loss.

## IV. Discussion and Conclusion

In this work, we presented a quantitative analysis framework for long-term, label-free characterization of live organoids using HT. By integrating a torus-shaped spatial filtering strategy, a multi-stage 3D segmentation pipeline, and a model-based dry-mass quantification approach, the proposed framework mitigates key consequences of the missing-cone artifact and enables reliable extraction of morphological and biophysical metrics from HT data. Application to expanding, collapsing, and fusing hepatic organoids demonstrated that the framework captures dynamic remodeling processes and reveals trends that are difficult to assess using conventional imaging and analysis methods.

A key strength of the proposed framework is its ability to quantitatively track organoid-scale physiological features across diverse perturbations. In expanding organoids, epithelial and lumen volumes increased in a coordinated manner, accompanied by stable surface-area ratios and nearly constant dry-mass density, indicative of preserved morphometric and biophysical homeostasis. In fusing organoids, transient deviations in morphological and biophysical parameters were followed by partial recovery, illustrating the framework's capacity to resolve non-monotonic and heterogeneous remodeling dynamics. These results highlight the potential of label-free HT combined with quantitative analysis for monitoring viability, assessing organoid-level robustness, and probing emergent homeostatic-like behaviors in 3D multicellular systems.

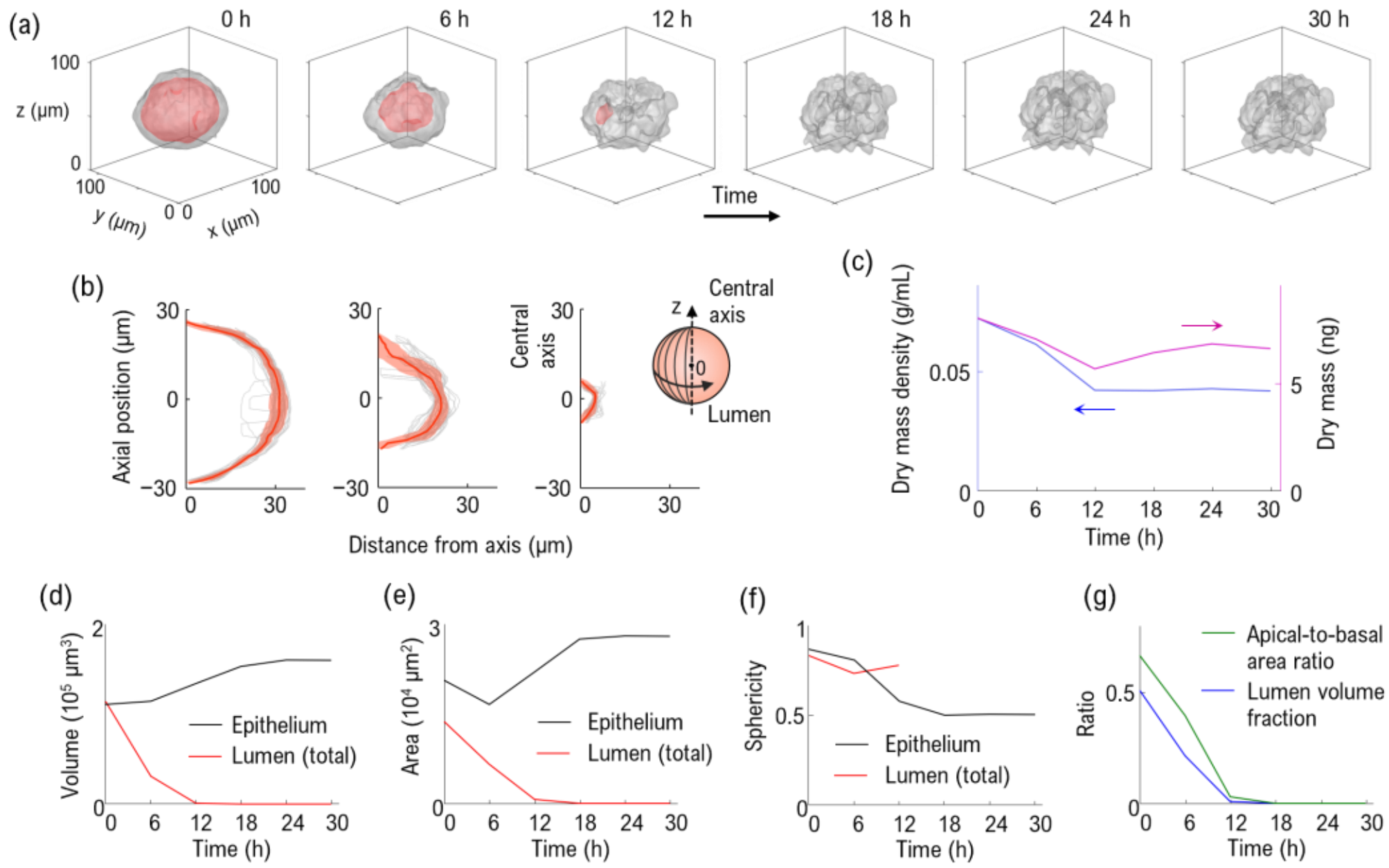

**Fig. 6. Quantitative analysis of a collapsing organoid over a 30-h time course.** (a) Three-dimensional renderings of the segmented epithelium and lumen at representative time points, illustrating progressive lumen collapse. (b) Lumen line profiles extracted along the vertical axis passing through the lumen centroid. Gray curves correspond to individual radial profiles sampled from the lumen surface at 10° angular intervals, while the red curve denotes their mean and the shaded region indicates the standard deviation. Using the segmented epithelium and lumen masks, temporal changes in (c) epithelial and lumen volumes, (d) epithelial and lumen surface areas, (e) lumen sphericity, and (f) the corresponding volume and surface-area ratios relative to the entire organoid were quantified. (g) Dry-mass density (blue) and total dry mass (purple) estimated from the epithelial region are plotted as functions of time using dual y-axes.

Although the results presented in this section focus on hepatic organoids as a representative application, the proposed MP-HT–based analysis framework is not specific to organoid systems. The underlying methodology relies on general properties of HT imaging and morphology-preserving representation, and is therefore readily applicable to a broad range of 3D biological specimens. In particular, the framework can be extended to other preclinical imaging contexts involving thick, heterogeneous samples, such as early-stage embryos [54], microfluidics [55]–[57], and thick tissue sections [58]–[60], where long-term, label-free volumetric imaging and robust morphological quantification are required. These potential extensions highlight the generality of the proposed approach beyond the specific organoid models demonstrated here.

Despite these advantages, several limitations inherent to current HT systems affect the accuracy of the proposed analysis. First, the RI reconstruction relies on the single-scattering approximation, which becomes increasingly invalid in deeper regions where multiply scattered light contributes to the measured signal [43], [44]. In our data, this limitation manifested as depth-dependent contrast degradation and incomplete delineation of epithelial boundaries, necessitating additional segmentation steps to repair holes or discontinuities in the epithelium mask. Multiple scattering also reduces effective RI contrast, potentially leading to underestimation of dry mass, an effect that is expected to become more pronounced for organoids with substantial axial thickness.

Second, although the morphology-preserving strategy enables robust segmentation at the organoid scale, it relies on high-spatial-frequency RI texture. Subcellular regions with intrinsically homogeneous composition or low molecular density—such as lipid-rich compartments or regions near focal adhesions—may be underrepresented in the filtered representation. While this limitation has minimal impact on organoid-level analyses, it may restrict applicability to single-cell or subcellular investigations.

Future work will focus on extending the framework toward more complete physical modeling of light–matter interactions in thick specimens, including integration with multiple light scattering reconstruction algorithms [61]–[63], which are expected to improve resolution and contrast at increased imaging depths. In parallel, incorporation of machine-learning–based segmentation approaches may further enhance robustness and reduce the need for manual review. Overall, the presented framework establishes a foundation for non-invasive, high-content analysis of organoid dynamics and provides a practical pathway toward quantitative, longitudinal studies in disease modeling, drug screening, and regenerative medicine.

## Appendix

### A. Spatial-frequency support of HT

Similar to other imaging modalities, the RI tomogram obtained by HT can be characterized by an effective PSF.

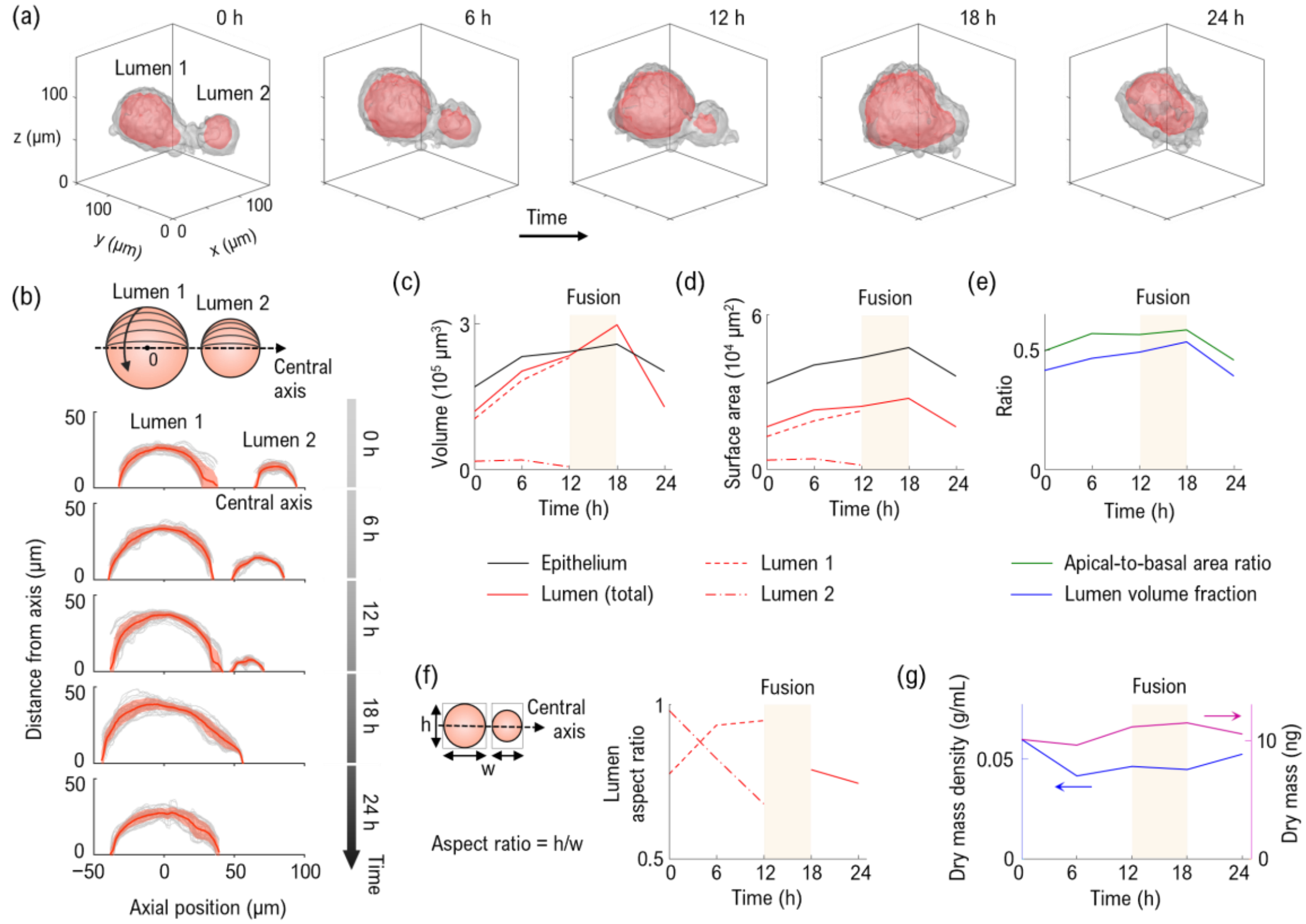

Fig. 7. Quantitative analysis of a fusing organoid over a 24-h time course. (a) Three-dimensional renderings of the segmented epithelium and lumen at representative time points, showing the progressive fusion of two initially separated lumens (lumen 1 and lumen 2). (b) Lumen line profiles extracted along the central axis connecting the two lumen centroids. Gray curves correspond to individual radial profiles sampled from the lumen surfaces at 10° angular intervals, while the red curve denotes their mean and the shaded region indicates the standard deviation. Using the segmented epithelium and lumen masks, temporal changes in (c) epithelial and lumen volumes, (d) epithelial and lumen surface areas, and (e) the corresponding volume and surface-area ratios relative to the entire organoid were quantified. (f) The lumen aspect ratio was computed from the line-profile analysis shown in (b). (g) Dry-mass density (blue) and total dry mass (purple) estimated from the epithelial region are plotted as functions of time using dual y-axes.

Unlike conventional imaging systems, HT inherently performs partial deconvolution during RI reconstruction. Consequently, the effective PSF is determined solely by the inverse Fourier transform of the system's spatial-frequency support over which the object spectrum is physically supported. This support region is defined by the autocorrelation of the Ewald spheres corresponding to the illumination and detection geometries. Each Ewald sphere represents the set of wave vectors of propagating light in the illumination or detection path and encodes the NAs of the optical system as well as the illumination wavelength.

In this study, both the illumination and detection NAs were set to 0.72, and the illumination wavelength was either 450 nm or 660 nm, depending on the imaging system. The spatial-frequency support was computed by numerically generating the corresponding Ewald shells in the 3D frequency domain, calculating their autocorrelation, and thresholding the result at 0.5. The resulting support map was then Fourier transformed to obtain the effective PSF of the HT system. The complete numerical procedure is summarized in Algorithm 1. Based on the computed support, the geometrical parameters of the torus-shaped filter were determined: $\alpha$ was defined as the lateral spatial frequency at which the support exhibits its maximum axial thickness, and $\beta$ as the difference between this frequency and the smallest lateral frequency at which the axial thickness decreases to half of its maximum value.

---

**Algorithm 1:** Numerical computation of the spatial-frequency support of HT

---

**INPUT:** 3D spatial-frequency grids ( $k_x$ , $k_y$ , $k_z$ ); NA of illumination and detection ( $\mathrm{NA_c}$ , $\mathrm{NA_o}$ ); Illumination wavelength ( $\lambda$ ); Wavenumber in the medium ( $k_0$ )

**OUTPUT:** Binary 3D spatial-frequency support mask (S)

*# Initialize empty 3D grids*

$E_c \leftarrow$ 3D array filled with 0

$E_o \leftarrow$ 3D array filled with 0

*# Construct the illumination Ewald shell*

For every lateral frequency pair ( $k_x, k_y$ ) satisfying the illumination NA condition,

$\sqrt{k_x^2 + k_y^2} < \dfrac{\mathrm{NA_c}}{\lambda}$ :

1) Compute the theoretical axial frequency:

$k_z^* = \sqrt{k_0^2 - k_x^2 - k_y^2}$

2) Find the value of $k_z$ on the predefined grid that is closest to $k_z^* - k_0$

---

3) Set the corresponding grid point in $E_c$ to 1, otherwise 0.

*# Construct the detection Ewald shell*
Repeat the above steps using the detection NA $NA_o$, producing the shell mask $E_o$.

*# Compute the autocorrelation of the two Ewald shells*
Use the convolution theorem:

$$R \leftarrow \left| \mathrm{IFFT}\left( \mathrm{FFT}(E_c)\, \mathrm{FFT}(E_o)^* \right) \right|$$

*# Threshold to obtain spatial-frequency support*
S ← R > 0.5

### B. Selection of the structuring element for excess-thickness erosion

To determine the optimal ellipsoidal structuring element for the erosion operation, we evaluated a family of ellipsoids with lateral radii $r_{x,y} \in \{1,2,\cdots,8\}$ pixels and an axial radius $r_z$= round($r_{x,y}$ / scalefactor). Here, the scalefactor represents the ratio of the axial pixel size to the lateral pixel size and ensures that the ellipsoidal structuring elements reflect the anisotropic voxel dimensions.

For each candidate radius, a 3D binary ellipsoidal mask was generated and applied to erode the phantom mask. The eroded phantom mask was then compared with the corresponding sample mask using an intersection-over-union (IoU) metric. The structuring element that yielded the highest IoU was selected as the optimal erosion element.

### C. Lumen-Epithelium Separation pipeline

**Algorithm 2:** Lumen-Epithelium Segmentation Pipeline

**INPUT:** Organoid mask (M); Axial scale factor (s); Maximum number of lumen centers (N)
**OUTPUT:** Lumen mask (L); Epithelium mask (E)

*# 1. Lumen Seed Detection*
1) Compute a 3D distance transform of the epithelium mask.
2) Detect regional maxima using extended maxima transform with threshold (t = 8).
3) Remove components touching the image border (26-connectivity).
4) Retain only components fully contained within the bounding box of M.
5) Sort components by volume and select the largest N = 3 seed centroids.

*# 2. Spherical Min–Max Contour Estimation for Initial Voxel Classification*
1) Extract the largest connected component of M.
2) Sample space with 50 azimuth × 50 elevation bins considering the axial scale factor.
3) Compute $R_{min}(\varphi,\theta)$ and $R_{max}(\varphi,\theta)$ for each angular bin.
4) Apply angular padding of 10 bins.
5) Interpolate missing samples linearly.
6) Apply median filters: 3×3 for $R_{max}$; 7×7 for $R_{min}$.
7) Remove padding.
8) Assign all voxels to spherical coordinates (r,φ,θ).
9) Mark voxels as lumen if $r < R_{min}(\varphi,\theta)$.
10) Mark extracellular if $r > R_{max}(\varphi,\theta)$.
11) Aggregate across seeds.
12) Remove voxels belonging to the original mask M.

*# 3. Dual-Sided Morphological Reconstruction*
1) Initialize lumen (L) and extracellular (BG) masks based on the previous voxel classifications.
2) Repeat
    Dilate L with 3×3×3 structuring element.
    Prevent L growing into BG or M.
    Mark collision voxels as boundary B.
    Dilate BG with 3×3×3 structuring element.
    Prevent BG growing into L or M.
    Mark collision voxels as B.
  Until L no longer grows.

*# 4. Final Mask Assembly*
E ← M OR B
L ← L